\documentclass[%
        submission,
       final,
        notitlepage,
        narroweqnarray,
        inline,
        ]{ieee}

\usepackage{amsmath}
\usepackage{amssymb}

\begin{document}

%----------------------------------------------------------------------
% Title Information, Abstract and Keywords
%----------------------------------------------------------------------
\title{Progress on indium and barium single ion optical frequency standards}

% format author this way for journal articles.
%\author[SHORT NAMES]{%
%      First Author\member{Fellow}
%      \authorinfo{%
%      A. Author is with the Department of Electrical Engineering,
%      Some University, Somewhere CA, 90210, USA,
%      Phone: \mbox{(xxx) xxx-xxxx}, email: \mbox{xxx@xxxx.xxx.xxx}}
%    \and
%      Second Author\member{Senior Member}
%      \authorinfo{%
%      B. Author is with the Department of Electrical Engineering...}
%    \and
%      and Third Author\member{Student Member}
%      \authorinfo{...}
%  }

% format author this way for conference proceedings
\author[J.A. Sherman, et al.]{%
      Jeff A. Sherman,%\member{}
      \authorinfo{Dept. of Physics\\
      University of Washington, Seattle, WA 98195\\
      Phone: (206) 685-0956, email: jeff.sherman@gmail.com}
    \and
      William Trimble,%\member{Senior Member}
    \and
      Steven Metz,%\member{Student Member}
     \and
      Warren Nagourney,%\member{}
      \and
      and Norval Fortson%\member{}
  }

% specifiy the journal name
%\journal{IEEE Transactions on Something, 1997}

% Or, when the paper is a preprint, try this...
%\journal{IEEE/LEOS Summer Topicals, San Diego, CA, USA, July 25--27, 2005}

% Or, specify the conference place and date.
%\confplacedate{San Diego, CA, USA, July 25--27, 2005}

% make the title
\maketitle               
\thispagestyle{empty}

% do the abstract
\begin{abstract}
We report progress on $^{115}$In$^+$ and $^{137}$Ba$^+$ single ion optical frequency standards using all solid-state sources.  Both are free from quadrupole field shifts and together enable a search for drift in fundamental constants.
\end{abstract}

% do the keywords
%\begin{keywords}
%keyword 1, keyword 2 ...
%\end{keywords}

% start the main text ...
%----------------------------------------------------------------------
% SECTION I: Introduction
%----------------------------------------------------------------------
\section{Introduction}

\PARstart In single ion optical frequency standards the narrow transitions of weakly allowed lines serve as high quality references largely free from shifts and perturbations such as the first-order Doppler effect.  Other research groups have demonstrated precise control or near-elimination of most effects, such as the second-order Doppler shift, quadratic Stark shifts, Zeeman shifts, and ac Stark shifts from blackbody radiation for a diverse set of ion species (In$^+$ \cite{Becker01}, Hg$^+$ \cite{Diddams01}, Yb$^+$ \cite{Blythe03,Tamm00}, Sr$^+$ \cite{Barwood99}, Ca$^+$ \cite{Champenois04}, and proposed in Tl$^+$ \cite{Torgerson00}).  However, one important effect that remains to be suitably controlled at the Hz level is the quadrupole Stark shift---that is, the shift arising from coupling of the trapped ion's atomic quadrupole moment to stray dc electric field gradients unavoidably present in the trap region.

Here we propose two single ion optical frequency standards that are inherently free from this effect and are under active development.  Solid-state lasers for all the required transitions are commercially available.  In addition to being viable candidates for absolute frequency standards, their construction can also yield a search for drift in fundamental physical constants.

\begin{figure}
\centering
\includegraphics[scale=0.5]{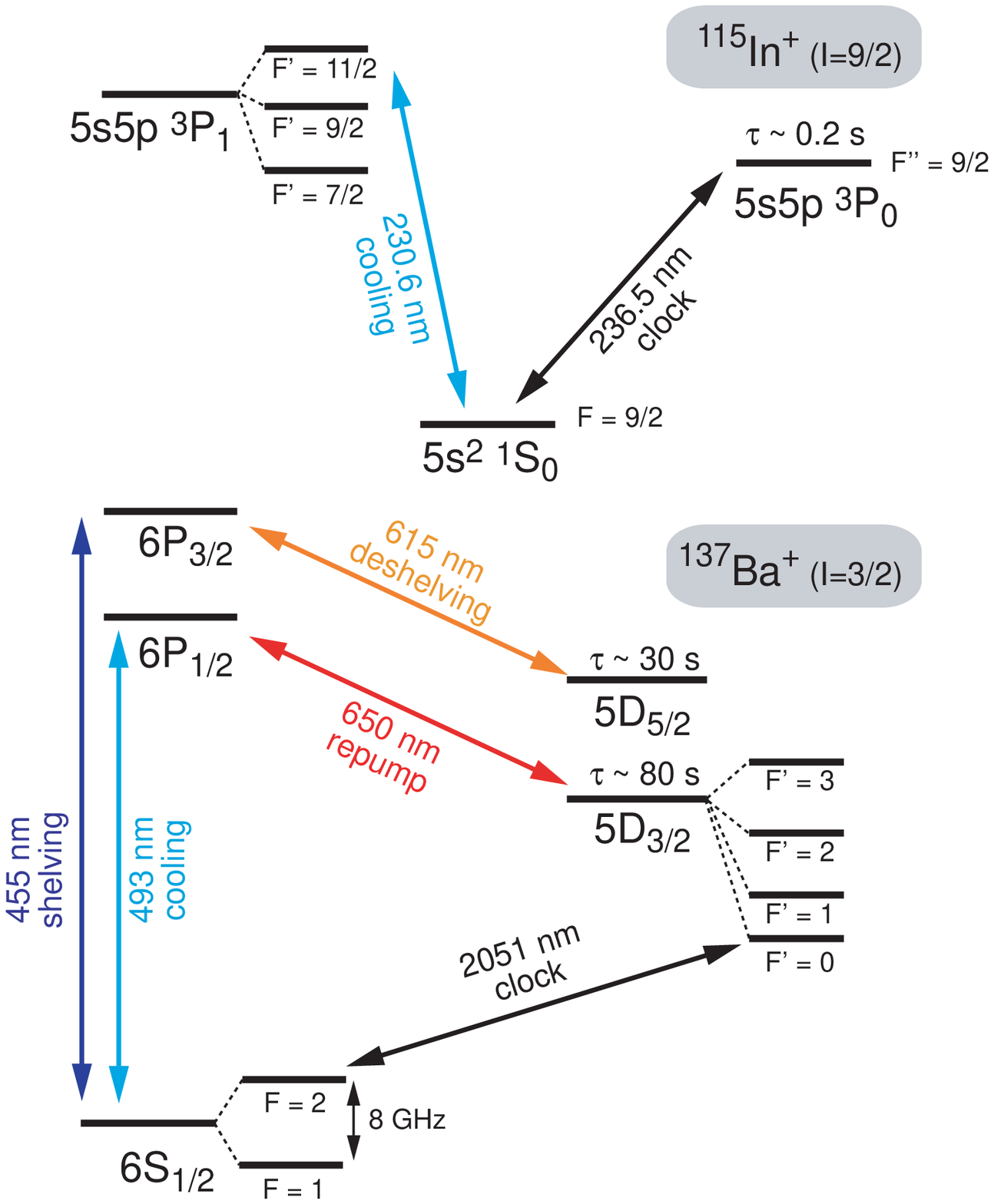}
\caption{Relevant atomic levels and transitions for $^{115}$In$^+$ and $^{137}$Ba$^+$  frequency standards.  All the wavelengths are accessible with solid state lasers as discussed in the text. 
%The direct 1.76~$\mu$m $6S_{1/2} \leftrightarrow 5D_{5/2}$ shelving transition for Ba$^+$ is not shown.
}
\end{figure}

\section{Indium ion frequency standard}
We trap single indium ions in a twisted wire Paul-Straubel trap with 700 volts of 10~MHz rf.  We then cool the ions to the Lamb-Dicke regime using 230.6~nm light resonant with the $^1S_0 \leftrightarrow {^3}P_1 (F = 11/2)$ transition.  This radiation is generated by frequency quadrupling a 922~nm external cavity diode laser/tapered amplifier system in resonant build-up cavities containing KNbO$_3$ and BBO nonlinear crystals.  The $^3P_0$ excited state is long-lived ($\tau = 195(8)$~ms \cite{Becker01}) giving the $^1S_0 \leftrightarrow {^3}P_0$ transition a natural linewidth of $\sim 1$~Hz and a quality factor of $Q = 1.5(10)^{15}$.  We create 236.5~nm clock light by frequency quadrupling the output of a monolithic Nd:YAG laser operating at 946~nm.  This laser, manufactured by InnoLight GmbH, is a non-planar ring oscillator with a nominal free-running linewidth of 2~kHz.  Similar laser systems have been narrowed below 1~Hz \cite{Webster03} by Pound-Drever-Hall locking to an isolated high finesse reference cavity.

Excitation of the clock transition is detected via the absence of cooling fluorescence when the ion is `shelved' in the metastable $^3P_0$ state.  A software servo system detects and corrects for long-term drifts in the clock laser while it is stabilized over short times using a stable reference cavity or other absolute reference, such as a microwave standard and femtosecond frequency comb.

An important feature of $^{115}$In$^+$ is its immunity to quadrupole Stark shifts due to both ground and excited clock states having zero electronic angular momentum $J = J' = 0$.  From Ref.\ \cite{Itano01}, quadrupole Stark shifts take the form
\begin{equation}
H_Q = \boldsymbol{\nabla} \mathbf{E}^{(2)} \cdot \boldsymbol{\Theta}^{(2)}
\end{equation}
where the tensor $\boldsymbol{\nabla} \mathbf{E}^{(2)}$ describes the electric field gradient at the site of the ion and $\boldsymbol{\Theta}^{(2)}$ are the set of irreducible second-rank tensors.  In the $IJ$-coupling approximation, shifts of the clock frequency scale with the reduced matrix elements
\begin{equation}
\langle n(IJ)F || \Theta^{(2)} || n (IJ) F \rangle \propto \left\{ \begin{matrix} J & 2 & J \\ F & I & F \end{matrix} \right\} \langle n J J | \Theta_0^{(2)} | n J J \rangle
\label{eq.quad}
\end{equation}
where the 6$j$-symbol and the atomic quadrupole moment vanish for $J=0$.

\section{Barium ion frequency standard}
Our barium ion trap \cite{Koerber03} operates at a similar frequency and strength as the indium trap but is constructed entirely from non-magnetic tungsten and tantalum.  Passive shielding has achieved $\le 10$ $\mu$G magnetic field stability measured using the ion \cite{Sherman05}.  We cool on the 493~nm $6S_{1/2} \leftrightarrow 6P_{3/2}$ transition using a frequency doubled 986~nm diode laser/tapered amplifier system while also applying a 650~nm diode laser which empties the long-lived $5D_{3/2}$ state.  Currently, a static magnetic field ($B \sim 1.6$ G) aligned along both lasers eliminates dark states \cite{Berkeland02} but for clock applications a low magnetic field is desirable; polarization modulation will accomplish dark state quenching \cite{Devoe02}.  High intensity filtered light-emitting diodes perform shelving to and deshelving from the $5D_{5/2}$ state and could ultimately be replaced by a 1.76~$\mu$m diode laser acting on the $6S_{1/2} \leftrightarrow 5D_{5/2}$ electric-quadrupole transition previously investigated as a clock candidate \cite{Yu94}.

The  Ba$^+$ $5D_{3/2}$ state lifetime ($\tau \sim 80$~s \cite{Yu97,Madej90}) is among the longest-lived of successfully cooled ions, giving the $6S_{1/2} \leftrightarrow 5D_{3/2}$ clock transition at 2051~nm a quality factor $Q = \nu / \Delta \nu > 10^{16}$. A diode pumped Tm,Ho:YLF solid state laser manufactured by CLR Photonics produces $> 80$~mW at 2051~nm with a free-running linewidth we have determined to be $\sim 10$~kHz.  We have detected the transition in $^{138}$Ba$^+$.

$^{137}$Ba$^{+}$ shares the advantage with In$^+$ of having vanishing shifts due to quadrupole Stark shifts.  The atomic quadrupole moment in Eq.~\ref{eq.quad} vanishes for the ground state $6S_{1/2} (F=2)$ and the $6j$-symbol in Eq.~\ref{eq.quad} vanishes for the excited clock state $5D_{3/2} (F=0)$.  Small shifts likely remain due to configuration mixing and coupling to the nuclear moment but these perturbations are expected to be much less than 1~mHz.

\renewcommand{\arraystretch}{1.25} 
\begin{table}
\centering
\begin{tabular}{l | c | c}
 &  $^{115}$In$^+$ & $^{137}$Ba$^+$ \\ \hline \hline
Clock frequency (cm$^{-1}$) & 42275.995  & 4873.720 \\
Clock state lifetime (s)  & 0.2  \cite{Becker01} & 80  \cite{Yu97} \\
Quality factor $\nu / \Delta \nu$ & $1.5(10)^{15}$ & $7(10)^{16}$ \\ \hline
Zeeman shift & 224 Hz/G \cite{Becker01} & -11 kHz/G$^2$ \\ \hline
\parbox{3.4 cm}{Clock drift (Hz/yr) \\ (if $\dot{\alpha}/\alpha = 10^{-15} \text{ yr}^{-1}$)} & 0.23 \cite{Angstmann04} & 0.35 \cite{Dzuba00} \\
\end{tabular}
\caption{Selected properties of the two single ion frequency standards discussed here.}
\label{tab.clock}
\end{table}

\section{A search for drift in fundamental constants}
Besides applications in navigation, long baseline interferometry, and detection of drift in pulsar periodicity, optical atomic frequency standards enable laboratory searches for drift in fundamental constants \cite{Bize03}.  Cosmological observations \cite{Murphy03} as well the analysis of the Oklo natural nuclear reactor \cite{Lamoreaux04} suggest drift in the fine-structure constant.  Since relativistic corrections to clock transition frequencies of different atomic species depend differently on the constant $\alpha$, a linear drift of $\alpha$ over time will be detected as an accumulated error between two or more atomic optical frequency standards.  Table \ref{tab.clock} shows the estimated clock frequency change that would occur in one year assuming a fractional change of $\dot{\alpha}/\alpha = 10^{-15}$ yr$^{-1}$ using latest theoretical results \cite{Angstmann04}, \cite{Dzuba00}.

% do the biliography:
\thispagestyle{empty}

\bibliographystyle{IEEEtran}
\bibliography{ionClocks}

\thispagestyle{empty}

\end{document}